# An open debate on SARS-CoV-2's proximal origin is long overdue

Rossana Segreto[1*], Yuri Deigin[2], Kevin McCairn[3], Alejandro Sousa[4,5], Dan Sirotkin[6], Karl Sirotkin[6], Jonathan J. Couey[7], Adrian Jones[8], Daoyu Zhang[9]

[1] Department of Microbiology, University of Innsbruck, Austria
[2] Youthereum Genetics Inc., Toronto, Ontario, Canada
[3] Synaptek - Deep Learning Solutions, Gifu, Japan
[4] Regional Hospital of Monforte, Lugo, Spain
[5] University of Santiago de Compostela, Spain
[6] Karl Sirotkin LLC, Lake Mary, FL 32746, USA
[7] University of Pittsburgh, School of Medicine, USA
[8] Independent bioinformatics researcher
[9] Independent genetics researcher

*Correspondence: rossana.segreto@uibk.ac.at

## Abstract

There is a near consensus view that SARS-CoV-2 has a natural zoonotic origin; however, several characteristics of SARS-CoV-2 taken together are not easily explained by a natural zoonotic origin hypothesis. These include: a low rate of evolution in the early phase of transmission; the lack of evidence of recombination events; a high pre-existing binding to human ACE2; a novel furin cleavage site insert; a flat glycan binding domain of the spike protein which conflicts with host evasion survival patterns exhibited by other coronaviruses, and high human and mouse peptide mimicry. Initial assumptions against a laboratory origin, by contrast, have remained unsubstantiated. Furthermore, over a year after the initial outbreak in Wuhan, there is still no clear evidence of zoonotic transfer from a bat or intermediate species. Given the immense social and economic impact of this pandemic, identifying the true origin of SARS-CoV-2 is fundamental to preventing future outbreaks. The search for SARS-CoV-2's origin should include an open and unbiased inquiry into a possible laboratory origin.

## Introduction

Severe acute respiratory syndrome coronavirus type 2 (SARS-CoV-2) is a novel Betacoronavirus of lineage B (subgenus Sarbecovirus) and the causative agent of Covid-19, the first detected cases of which were identified in Wuhan in December 2019 (Huang et al., 2020). The near-consensus view of the origin of SARS-CoV-2 is a natural zoonosis (Zhu N. et al., 2020, Wu, F. et al., 2020, Zhou P. et al., 2020). Bats are thought to be the natural reservoir for SARS-related coronaviruses (SARS-r CoVs) (Li et al., 2005; Wang et al., 2006) and have been identified as the ancestral source from which severe acute respiratory syndrome coronavirus (SARS-CoV) evolved (Janies, 2008; Sheahan, 2008). While several intermediate host species have been proposed as the zoonotic source for SARS-CoV-2 (Xiao et al., 2020; Lam et al., 2020; Zhang, T. et al., 2020; Zhou and Shi, 2020), the source of direct bat to human, or intermediate animal to human zoonotic transmission of SARS-CoV-2 has not been established. An alternative hypothesis, that SARS-CoV-2 leaked from a laboratory, has been widely dismissed (Rasmussen, 2020), yet very few papers counter this theory with data analysis (Andersen et al., 2020; Liu SL. et al., 2020; Graham and Baric, 2020).

Here, we address the main arguments in support of a natural origin of SARS-CoV-2, and outline the various points which support the alternative, that a laboratory origin is still a valid possibility that should not be discounted. To help prevent future viral pandemics, it is of pivotal importance to identify the source of the virus and this is only possible with an unbiased analysis of all data available. We couple this work with calls from recent opinion pieces and comparative studies questioning a zoonotic origin (Sousa, 2020; Sirotkin & Sirotkin 2020; Relman, 2020; Segreto & Deigin 2020; Butler, 2020; Sallard et al., 2021) via a review of the latest literature, and propose an alternative to the natural zoonosis hypothesis.

## Early outbreak and pangolins

The earliest detected cases of COVID-19 were located in Wuhan, China (Huang et al., 2020). The Huanan seafood market in Wuhan had initially been posited as a possible location of initial zoonotic transfer from wild animals to humans (Huang et al., 2020). However, three of the four patients with the earliest recorded onset of COVID-19 symptoms, none had an association with the seafood market (Huang et al., 2020), and the ancestral T8782 and C28144 genotype was not associated with the seafood market (Chen et al., 2020). Phyloepidemiologic analysis of early cases (Yu et al., 2020) also discounted this theory.

Although the exact zoonotic agent of the original SARS-CoV has not been identified, Chiroptera are considered to be the natural reservoir of SARS-r CoVs (Li et al., 2005). Initially, palm civets and raccoon dogs were proposed as zoonotic agents (Chinese SARS Molecular Epidemiology Consortium, 2004) or intermediate hosts (Tang et al., 2006), but it remains possible that they were themselves infected by humans instead (Janies et al. 2008).

For SARS-CoV-2, several authors have proposed pangolins as an intermediate host due to the similarity of the RBD for pangolin-CoV to SARS-CoV-2 RBD (Xiao et al., 2020; Lam et al., 2020; Zhang, T. et al., 2020). Pangolins are, however, unlikely to be the intermediate host for SARS-CoV-2.  Although two pangolin-CoV's (Xiao et al. 2020, Liu, P. et al., 2020) exhibited strong binding to human ACE2,  binding to pangolin ACE2 was approximately tenfold weaker, and binding to bat Rhinolophus ferremequinum ACE2 was very weak, which is comparable to that exhibited by SARS-CoV-2 (Wrobel et al. 2021). This indicates that neither pangolin-CoV had adapted well to pangolins and that more research is required to validate the viability of coronaviruses to spread naturally between pangolins. Because of a 10-15% divergence throughout the entire spike protein, with the exclusion of the N-terminal domain, Boni et al. (2020) conclude that SARS-CoV-2 is unlikely to be a recombinant of an ancestor of pangolin-CoV and the closest SARS-CoV-2 relative, RaTG13.

All published pangolin-CoV genome sequences with a nearly identical spike RBD to SARS-CoV-2 were sourced from a single batch of smuggled pangolins (Chan and Zhan 2020), raising the question whether pangolins may have been infected from another host species or from humans during trafficking (Choo et al. 2020; Wenzel, 2020). Unlike other species demonstrated to be vectors for CoVs, pangolins are not trafficked together live caged in large groups for extended periods of time, making this an unlikely scenario for viral enhancement. Also, pangolins are critically endangered (Choo et al., 2020), exhibit a solitary nature, potentially have limited infection resistance (Choo et al. 2016), and a recent screening of 334 pangolins revealed a lack of CoV infections of pangolins in the wild (Lee et al., 2020). Finally, the discovery of synthetic DNA sequences in pangolin CoV metagenomic raw sequence reads by Zhang, D., (2020) and the interpretation that the pangolin-CoV genomes were generated from a synthetic construct, requires further investigation.

## Evolution

Unlike SARS-CoV in its early and middle phases (Chinese SARS Molecular Epidemiology Consortium 2004; Sheahan et al., 2008, Janies et al., 2008), or the evolution of middle east respiratory syndrome-related coronavirus (MERS-CoV) (Lau et al., 2017, Forni et al., 2017), SARS-CoV-2  exhibits limited diversity across its genomes (Dearlove et al., 2020, van Dorp et al., 2020, Zhan et al., 2020, Jia et al., 2020). A very recent emergence of SARS-CoV-2 into the human population has been proposed based on the sampling of eight nearly identical complete genomes in December 2019 (Lu et al., 2020). From earliest strains in Wuhan in 2019, SARS-CoV-2 resembled SARS-CoV in the late phase of its 2003 epidemic after SARS-CoV had developed several advantageous adaptations for human transmission (Zhan et al., 2020).

While there is no record of a process of early evolutionary adaptation, SARS-CoV-2's RBD appears to be highly optimized for binding to human ACE2 (Delgrado Blanco et al., 2020; Damas et al., 2020). In this respect, 43% of modelled mutations destabilize the binding energy of the SARS-CoV-2 spike protein RBD to hACE2, while just 1% of the mutations stabilize it (Delgrado Blanco et al., 2020). Substitution of any of the eight SARS-CoV-2 RBD residues proximal to the hACE2 binding interface with the residues found in the RaTG13 RBD were shown to be detrimental to hACE2 binding (Conceicao et al., 2020). Furthermore, Piplani et al., (2020) in a study of 13 animal species including Manis javanica and Rhinolophus sinicus found that the SARS-CoV-2 spike protein had the highest overall binding energy for human ACE2.

Because bats are considered to be natural reservoirs of SARS-r CoVs (Li et al., 2005), SARS-CoV-2 to ACE2 binding ability is expected to be high in bats. However bat species are poorly infected by SARS-CoV-2 and they are therefore unlikely to be the direct source for human infection. SARS-CoV-2 does not replicate in R. sinicus kidney or lung cells (Chu et al., 2020), binds poorly to R. sinicus ACE2 (Tang, Y. et al., 2020; Li, Y. et al., 2020; Piplani et al., 2020) and exhibits no binding to R. ferrumequinum ACE2 (Tang, Y. et al., 2020). In addition, in-silico modelling of the binding affinity for 37 bat species (Damas et al., 2020) showed that eight species exhibited very low binding affinity (including R. ferrumequinum), and the other 29 exhibited low (including R. pearsonii and R. sinicus) binding ability. Although the host of RaTG13, R. affinis was not modelled by Damas et al. (2020), these results are perplexing as it indicates a significant and unexplained evolutionary distance between SARS-CoV-2 and bats. Curiously RaTG13 also exhibits poor binding to R. sinicus ACE2 (Li, Y. et al. 2020), R. pusillus ACE2 (Chu et al. 2020), R. ferremequinum (Wrobel et al. 2021) and hACE2 (Wrobel et al., 2020, Wrobel et al. 2021). The combination of high human adaptation and poor bat susceptibility from the first sampled strains of SARS-CoV-2 differs greatly from the evolution of MERS-CoV and SARS-CoV.

Recombination processes have been proposed by several authors as a mechanism by which SARS-CoV-2 may have evolved. Interestingly, there is no evidence of recombination events in studies of SARS-CoV-2 by Richard et al. (2020) (6,546 genome sequences as of September 2020) or Bobay et al. (2020) (218 sequences as of August 2020). This is in contrast with MERS-CoV, where despite a much smaller sample size, recombination events were detected. Furthermore, there is also no indication of recombination between the subgenus Sarbecorvirus and other Betacoronavirus subgena or species of the Alpha, Gamma or Deltacoronavirus genera. Indeed, in the subgenera of Betacoronaviruses: Embecovirus, Merbecovirus and Sarbecovirus, gene exchange is restricted to members of the same subgroup (Bobay et al., 2020). The hypothesis that the RBD of the SARS-CoV-2 spike protein arose via a recent recombination with a pangolin hosted CoV RBD (Andersen et al., 2020; Xiao et al., 2020; Li, X. et al., 2020; Lam et al., 2020; Zhang, T. et al., 2020) is not likely (Bobay et al., 2020; Paraskevis et al. 2020) and poor taxon sampling by Zhang, T. et al. (2020), Lam et al. (2020) and Xiao et al. (2020) is discussed by Wenzel (2020). Although earlier recombination and mutations have been proposed (Bobay et al., 2020; Wang, H. et al., 2020; Patiño-Galindo et al., 2020), given that Sarbecoviruses have not been shown to recombine with other CoV genera, or other Betacoronavirus subgena, the acquisition of an RBD or a novel furin cleavage site (FCS) insert by SARS-CoV-2 (Tang, T. et al., 2020) is not likely to have happened through this natural mechanism. The hypothesis of Gallaher (2020) that SARS-CoV-2's FCS might have resulted from a recombination event of a RaTG13-like CoV and HKU-9, which is a lineage D Betacoronavirus is also unlikely to be valid, especially in light of RaTG13 being hosted by mircobats (Rhinolophus genus) and HKU-9 by megabats (Rousettus genus).

## Furin cleavage site

SARS-CoV-2 is the only Sarbecovirus to contain a FCS (Coutard et al., 2020). Indeed, no CoV with a spike protein sequence homology of greater than 40% to SARS-CoV-2 has a FCS (Wu, C. et al., 2020). The multibasic FCS (Fig. 1) ('RRAR↓', the arrow indicates site of proteolytic cleavage) in SARS-CoV-2 plays a key role in its pathogenesis (Johnson et al., 2020; Hoffman et al., 2020; Shokeen et al., 2020; Qiao and Olvera de la Cruz 2020; Lau et al., 2020; Shang et al., 2020) and enhances its human pathogenicity over a minimal FCS 'RXXR↓' (Thomas, 2002). It is also unusual, diverging from the canonical 'RX[K/R]R' motif (Tang, T. et al., 2020). The presence of an arginine at the third position P3 before the FCS increases the efficiency of the FCS tenfold (Henrich et al., 2003). Its presence is also rare, occurring in only 5 out of 132 known FCSs (Lemmin et al., 2020). The 'RRAR' motif conforms to the '[R/K]XX[R/K]' 'C-end rule', creating a binding site for cell surface neuropilin (NRP1 and NRP2) receptors (Teesalu et al., 2009), which are more widely expressed than ACE2. NRP1 has been demonstrated as an alternate route for virus entry (Cantuti-Castelvetri et al., 2020; Daly et al., 2020).

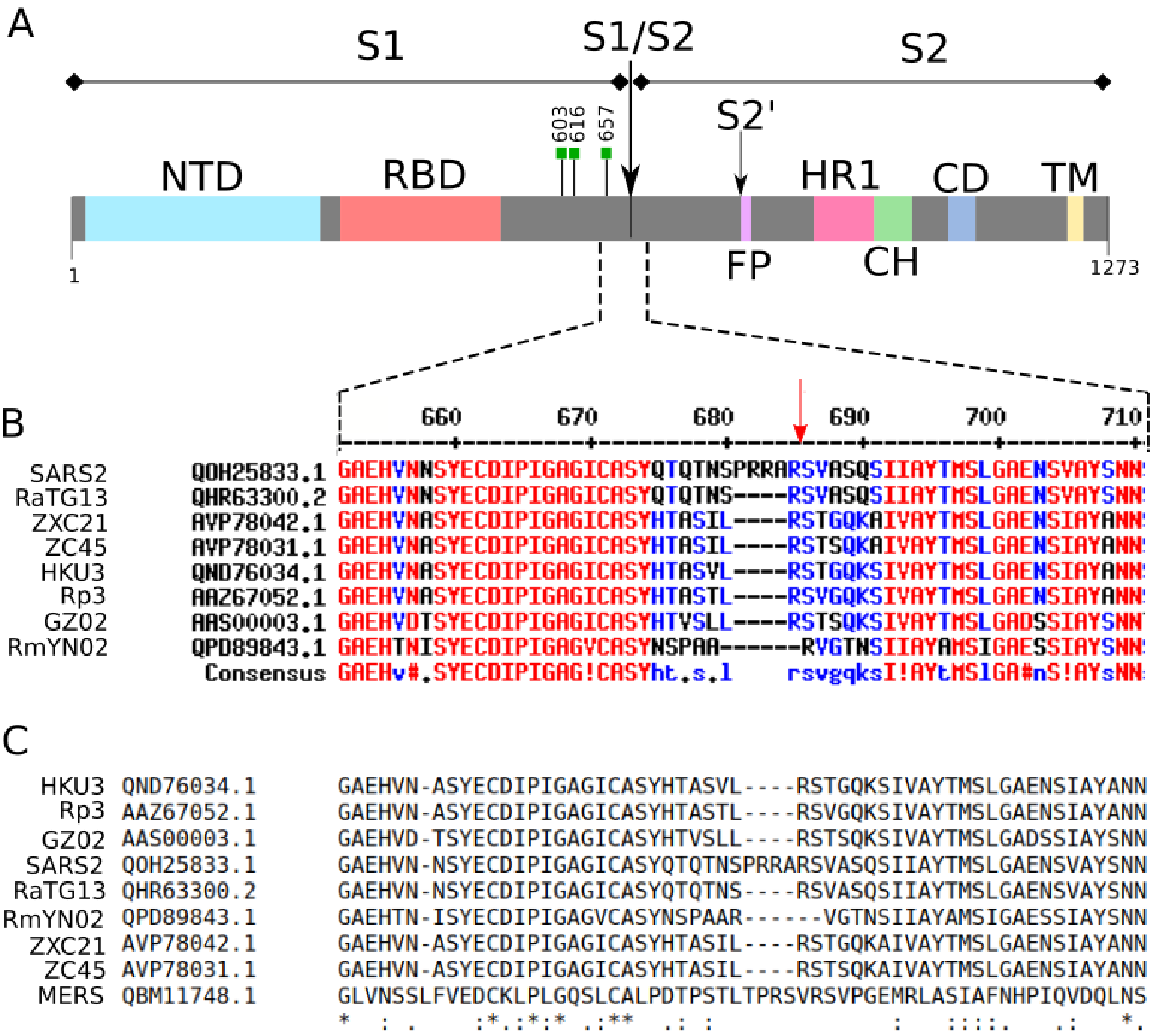

**Fig. 1.** Schematic representation of the SARS-CoV-2's spike protein showing subunits and domains as well as local sequence alignments with other *Betacoronaviruses*. A) Spike domains: N terminal domain (NTD); Receptor binding domain (RBD); Fusion peptide (FP); Hetapad repeat 1 (HR1); Central helix (CH); Connector domain (CD), Transmembrane domain (TM). The S1↓S2 and S2′ cleavage sites are indicated by arrows. Green boxes indicate location of N-glycosylated residues proximal to FCS. B) MultAlin alignments using SARS-CoV-2 spike protein sequence numbering reference (Corpet 1988). (http://www.sacs.ucsf.edu/cgi-bin/multalin.py). C) SACS ClustalW multiple sequence alignment (http://www.sacs.ucsf.edu/cgi-bin/clustalw.py). Definition and accession numbers as follows: SARS2: SARS-CoV-2 Wuhan-Hu-1 (QOH25833.1); RaTG13 (QHR63300.2); ZXC21: bat-SL-CoVZXC21 (AVP78042.1); ZC45: bat-SL-CoVZC45 (AVP78031.1); HKU3: Bat SARS coronavirus HKU3 (QND76034.1); Rp3: Rp3/2004 (AAZ67052.1); GZ02: SARS coronavirus GZ02 (AAS00003.1); RmYN02: Bat coronavirus RmYN02 (QPD89843.1); MERS: middle East respiratory syndrome-related coronavirus (QBM11748.1). SARS-CoV-2 referenced sequence indexes shown.

Because of the insertion of the FCS, not only furin, but also several other proteolytic enzymes are able to activate SARS-CoV-2's spike protein (Jaimes et al., 2020). The proline residue at position P5 (5th residue prior to the FCS) is rare and only appears in 5 out of 132 known FCSs (Lemmin et al., 2020). Proline has a restricted phi angular range in peptide bond formation (Morgan and Rubenstein, 2013) which imposes conformal restraints on the peptide chain and results in the separation of the cleavage site from other structural elements, facilitating exposure to host proteases (Lemmin et al., 2020). In comparison with 132 known FCSs in FurinDB (Tian et al., 2011) (http://www.nuolan.net/substrates.html), SARS-CoV-2's FCS exhibits several intriguing features. The P11-P1 'QTQTNSPRRAR' motif is homologous to neurotoxins from Ophiophagus and Bungarus genera and neurotoxin-like regions from RABV strains, and may act as a superantigenic fragment (Cheng et al., 2020). A lack of a basic R or K at the P2 position is shared by only 30 of the FCSs; the 'XXRR[A/S/C/G/T/V/I/L]R↓' motif is shared by only two other FCSs; and the 'XPXRXX↓' motif also only occurs in one other FCS in FurinDB. The 'XXRRAR↓XX' was found to be only shared by the bacterial toxin proaerolysin (in FurinDB) (Abrami et al., 1998) and Alphacoronavirus AcCoV-JC34 (Ge et al., 2017).

Another unique feature of SARS-CoV-2 when compared to related CoVs is a longer loop containing the S1/S2 cleavage site (Lemmin et al., 2020): it is at least 4 amino acids longer around the site containing the FCS than any other known Sarbecovirus. The combination of FCS and extended loop length facilitate SARS-CoV-2 activation by protease TMPRSS13, as well as TMPRSS2 albeit at one third the effectiveness of TMPRSS2 (Laporte et al., 2020). Mutants with a deleted 'PRRA' insert or a shortened FCS loop with deleted preceding amino acids 'QTQTN' abrogated the effectiveness of both TMPRSS2 and TMPRSS13 for facilitating cleavage (Laporte et al., 2020).

Moreover, the FCS in SARS-CoV-2 is coded by rare codons, leaving it not in-frame with the rest of the sequence; thus violating the rules of the copy choice recombination mechanisms that postulate in-frame insertions. Additionally, its insertion causes a peculiar split of one of the codons, serine (TCA) when compared with the close relatives MP789 and RaTG13 (Segreto & Deigin, 2020). The recent acquisition of the FCS by SARS-CoV-2 via a natural insert was proposed by Wu and Zhao (2020) on the basis of the existence of FCS in other, not closely related Betacoronaviruses with different loop positions to SARS-CoV-2 and the existence of a partial natural insert in the same region in RmYN02 (Zhou, H. et al., 2020). The reliability of the conclusions of Zhou, H. et al. (2020) has been questioned by Deigin and Segreto (2020), who particularly challenge the claim that RmYN02 has an insertion around the site of the FCS insertion in SARS-CoV-2 and instead point to a two amino acid deletion in RmYN02 at that locus. Therefore, RmYN02 should not be used as evidence of the natural origin of SARS-CoV-2's FCS until its claimed insertion is properly validated.

In several viruses, low affinity attachment to heparan sulphate (HS) improves the chances of binding to a more specific entry receptor by increasing viral concentration at the cell surface (Schneider-Schaulies, 2000; Zhu et al., 2011, Cagno et al., 2019). The binding to HS or allied polysaccharide heparin by SARS-CoV-2 has been demonstrated by several studies (Mycroft-West et al., 2020; Kim et al., 2020; Zhang, Q. et al., 2020; Clausen et al., 2020; Tiwari et al., 2020, Kwon et al., 2020) and heparin binding affinity in SARS-CoV-2 is much higher than in SARS-CoV or MERS-CoV (Kim et al., 2020). We note that the SARS-CoV-2 FCS 'PRRAR↓S' motif in its uncleaved state is consistent with the HS binding region motif 'XBBXBX' (where B is a basic and X is a hydropathic residue), one of two consensus motifs determined by Cardin and Weintraub (1989) by comparing several potential heparin-binding sites in selected proteins. This particular site in the FCS was demonstrated to have the highest HS binding affinity among the three glycosaminoglycan binding motifs identified in the spike protein of SARS-CoV-2 by Kim et al. (2020). Cell-culture adaptation to 'Cardin-Weintraub' motifs has been demonstrated in multiple cell passage studies (de Haan et al., 2005; de Haan et al., 2008, Millet et al., 2020) and it should be considered as a possible reason for the strong HS binding affinity identified in SARS-CoV-2. Indeed, while there is no significant O-glycosylation on the spike protein in human cells (Wang, D. et al., 2020), the use of insect cell culture and baculovirus display system, where glycoprotein sialylation is not a major biochemical process (Marchal et al., 2001) could allow O-glycosyltransferases access to the FCS. While in human cells we interpret this region to be shielded by N-linked complex large glycans N616 and N657 and mixed Oligomannose/complex type glycans at site N603 (Casalino et al. 2020, Sun et al., 2020, Watanabe et al., 2020), in insect cells we predict that O-glycosylation on S685 would prevent cleavage by furin and preferentially bind heparan sulphate (HS). Repeated passage through an insect cell culture from an inserted more potent artificial FCS motif could then lead to the generation of a cell-culture adaptive O-glycosylated 'RRAR' signature.

A minimal FCS could potentially have evolved via a single point mutation T678R (Li, W. et al., 2015), which is evolutionarily more parsimonious than a complete 12nt insertion of 'PRRA'. A multi basic cleavage site is also plausible with an additional mutation N679R. We note that deletions but not insertions frequently happen at the S1/S2 junction of SARS-CoV-2 during serial cell passage (Peacock et al., 2020), and have also been detected in strains isolated from hamsters and humans (Lau S-Y. et al. 2020; Liu Z. et al. 2020). The acquisition of the FCS via a natural insert, when a FCS could have evolved far more easily though point mutation, we believe is highly unlikely.

Because the presence and coding sequence of a FCS are important for pathogenesis, host range, and cell tropism (Nagai el al., 1993; Millet et al., 2015), the addition of a FCS into viruses has been an active area of gain-of-function research. A FCS can be easily inserted using seamless technology (Yount et. al., 2002; Sirotkin and Sirotkin 2020) without any need for cell passage, as previously performed in experiments on virulence and host tropism (Cheng et al. 2019). Insertions to change the properties of SARS-r CoV viruses are documented by Ren et al. (2008) and Wang et al. (2008). Considering that natural mutations have a very low probability to result in a stretch of 12 amino acids coding for an optimized FCS without any known intermediate form in Sarbecovirus, an artificial insertion of the FCS in SARS-CoV-2 may provide a more parsimonious explanation for its presence than natural evolution.

## Binding domains and peptide mimicry

A ‘ganglioside-binding domain’ (GBD) in the N terminal domain (NTD) of SARS-CoV-2 (Pirone et al., 2020) is characterized by a large flat interface enriched in aromatic and basic amino acid residues (Fantini et al. 2020a) and contains one of three inserts in the NTD of SARS-CoV-2 identified by Zhou P. et al. (2020a). The GBD proffers SARS-CoV-2 with a dual receptor/attachment ability to sialic-acid-containing glycoproteins, in addition to the primary ACE2 receptor. The importance of the GBD in SARS-CoV-2 infectivity was indicated by Chi et al. (2020) and McCallum et al. (2021) who identify potent binding antibodies which provide strong neutralizing activity against SARS-CoV-2 by binding to residues in this domain. Fantini et al. (2020b) discuss the flat structural topography of the GBD which proffers improved functional interaction, and because of this attribute and sequence peculiarities in the spike protein, raise questions concerning the proximal origin of SARS-CoV-2. The flat topography of the GBD was also observed by Seyran et al. 2021 as anomalous compared with other human CoVs, which typically exhibit hidden sugar-binding site localization as an evolutionary measure to evade host glycan-binding immune receptors, the “Canyon Hypothesis” (Rossman 1989; Chen and Li 2013; Li, 2015).

Another curious feature of SARS-CoV-2 is its binding efficiency to hACE2, being much more effective than SARS-CoV. Khatri et al. (2020) measured a large interaction surface with high binding-affinity between SARS-CoV-2 and ACE2 as >15-fold stronger than between SARS-CoV and ACE2. This is supported by Wrapp et al. (2020) who find ~10- to 20-fold higher binding efficiency. The increased SARS-CoV-2 to ACE2 binding efficiency has been proposed to be due to a larger hydrophobic interaction surface for SARS-CoV-2 over SARS-CoV (Gussow et al., 2020; Wan et al., 2020; Lai et al., 2020; Khatri et al., 2020, Brielle et al., 2020) with an increased number interacting residues (Brielle et al., 2020; Wang, Q. et al., 2020) and extra charge interaction (Sørensen et al., 2020; Gussow et al., 2020; Wang Y. et al., 2020). Closer interaction distances between the N-terminal end of ACE2 and the central region of the Receptor Binding Motif (RBM) for SARS-CoV-2 over SARS-CoV (Wang, Y. et al., 2020) also facilitates coupling. These modifications indicate a more highly adapted ability for SARS-CoV-2 to bind to ACE2 than seen for SARS-CoV. While SARS-CoV to human ACE2 affinity relied on five key residues all of which exhibited natural mutation in the early stages of adaptation to a new host (Wan et al., 2020), SARS-CoV-2 displays from even the very first isolates, a more optimized configuration without any evidence of early natural mutations (Zhan et al., 2020).

Other indications of significant human adaptation are seen in peptide mimicry by SARS-CoV-2. Sørensen et al., (2020) find that the SARS-CoV-2 spike protein is remarkably well adapted to humans with a 78.4% similarity to human epitopes. This finding is consistent with work by Kanduc and Shoenfeld (2020) who observe a highly improbable massive hexa and heptapeptide sharing between SARS-CoV-2 spike glycoprotein and human and Mus musculus proteins. Interestingly, mouse ACE2 does not effectively bind to the SARS-CoV-2's spike protein (Li, Y. et al., 2020; Tang, Y. et al., 2020; Praharaj et al., 2020; Damas et al., 2020). Extensive passage in mice with humanized lungs and immune systems (Cockrell et al., 2018) could explain such an improbable peptide sharing.

## O-linked glycans

Theoretical predictions for O-linked glycans in the SARS-CoV-2 spike protein have been used as evidence of a 'mucin-like domain' that might be involved in immunoevasion by shielding epitopes or key residues on the SARS-CoV-2 spike protein (Andersen et al., 2020), and hence supporting the argument for natural evolution of SARS-CoV-2.

O-glycosylation and/or N-linked sulfated glycans on full length SARS-CoV-2 spike protein constructs (Zhao et al., 2020; Watanabe et al., 2020; Klein and Zaia 2020; Sanda et al., 2020) and subunits (Shajahan et al., 2020) has been reported by several groups, albeit at relatively low levels of site occupation. Wang, D. et al. (2020) however in a comprehensive, high-fidelity mass spectrometric approach based on glycan reporter Signature Ions-Triggered Electron-Transfer/ Higher-Energy Collisional Dissociation (EThcD) Mass Spectrometry, did not observe any detectable occupied O-glycosylation sites. The use of EThcD allowed the sites of glycosylation to be unambiguously determined with a greater proportion of fragment ions observed (Riley et al., 2020). This method provides an increased degree of confidence in the results over conventional collision-induced dissociation, higher-energy collisional dissociation (HCD) (Watanabe et al., 2020; Zhang, Y. et al., 2020; Shajahan et al., 2020; Klein and Zaia, 2020), stepped collision energy HCD (Zhao et al., 2020), or HCD fragmentation and modulated normalized collision energy (Sanda et al., 2020) methods.

Furthermore glycan sequons can actually arise in vitro in the presence of antibodies, as was recently observed during serial passaging of SARS-CoV-2 (Andreano et al. 2020) and may also arise in the lab during in vivo passaging of viruses in, for example, humanized mice.

Critically, contrary to Andersen et al. 2020 supposition, there is no O-linked glycosylation on the neighboring residues of the S1/S2 junction, or at a significant level anywhere along the spike protein. No interaction of SARS-CoV-2 with a host immune system based on O-linked glycans can be claimed, and hence does not support the argument for natural evolution of SARS-CoV-2.

## Reverse-genetic systems and virus backbone

The observation that SARS-CoV-2 was not derived from a previously used virus backbone was used as an argument by Andersen et al. and Liu et al. 2020 as evidence against a laboratory origin hypothesis. The Betacoronavirus RaTG13 was fully sequenced in 2018 (Zhou P. et al., 2020b) but only published after the beginning of the pandemic (Zhou P. et al., 2020a). More unpublished sequences existed in a WIV database that was deleted after the beginning of the pandemic (Segreto & Deigin, 2020). SARS-CoV-2 could have been engineered using one of the over 1,500 strains openly collected by institutions associated with the WIV (Sirotkin & Sirotkin, 2020), a completely undocumented backbone, or one of several fairly well correlated bat-CoV's could have been used in combination with directed evolution, a widely used technique for introducing mutations and selection to achieve proteins with desired properties (Badran and Liu, 2015; Standage-Beier and Wang 2006; Simon et al., 2019). Specifically, this technique has been used for engineering novel virus variants (Excoffon et al., 2009; Lin et al., 2012; Meister et al., 2019). Furthermore, novel yet undocumented reverse-genetic systems could also have potentially been used. Indeed, multiple groups have developed SARS-CoV-2 reverse genetics systems for SARS-CoV-2 research in short periods of time (Hou et al., 2020; Torii et al., 2020; Thi Nhu Thao et al. 2020). Additionally, seamless "No See'em" technology pioneered nearly 20 years ago, allows reverse engineering to be used without leaving any traces (Yount et al., 2002).

We disagree with the hypothesis by Andersen et al. (2020) that the high-affinity binding solution of SARS-CoV-2's RBD to hACE2, which differs from the optimal binding solution modelled for SARS-CoV (Wan et al. 2020) provides strong evidence that SARS-CoV-2 could not have been engineered in a laboratory. Computational prediction is not necessary for generating novel human pathogenic viruses. Culturing and adapting CoVs and influenza A virus to different cell lines, including human airway epithelial cells, has been conducted in various laboratories (Tse et al., 2014; Menachery et al., 2015; Zeng et al., 2016; Jiang et al., 2020); furthermore, experimental creation of chimeric viruses by directed engineering as discussed above does not require prior modelling.

## Conclusion

More than a year after the initial documented cases in Wuhan, the source of SARS-CoV-2 has yet to be identified and the search for a direct or intermediate host in nature has been so far unsuccessful. The low binding affinity of SARS-CoV-2 to bat ACE2 studied to date, does not support Chiroptera as a direct zoonotic agent. Furthermore, the reliance on pangolin CoV RBD similarity to SARS-CoV-2 as evidence for natural zoonotic spillover is flawed as pangolins are unlikely to play a role in SARS-CoV-2's origin and recombination is not supported by recent analysis. At the same time, genomic analyses pointed out that SARS-CoV-2 exhibits multiple peculiar characteristics not found in other Sarbecoviruses. A novel multibasic FCS confers numerous pathogenetically advantageous capabilities, the existence of which is difficult to explain though natural evolution; SARS-CoV-2 to hACE2 binding is far stronger than SARS-CoV, yet there is no indication of amount of evolutionary adaptation that SARS-CoV or MERS-CoV underwent. The flat topography of the GBD in the NTD of SARS-CoV-2 does not conform with typical host evasion evolutionary measures exhibited by other human CoVs. The combination of peptide mimicry to humans and mice, physical structure and binding strength, as well as high adaptation for human infection and transmission from the earliest strains might suggest the use of humanized mice for the development of SARS-CoV-2 in a laboratory environment. The application of mouse strains expressing hACE2 for SARS-CoV related research is well documented (Ren et al., 2008, Hou et al., 2010, Menachery et al., 2015, Cockrell et al., 2018, Jiang et al., 2020). Additionally, culturing and adapting CoVs to different cell lines, including human airway epithelial cells has been experimentally conducted in various laboratories (Tse et al., 2014; Menachery et al., 2015; Zeng et al., 2016; Jiang et al., 2020). While a natural origin is still possible and the search for a potential host in nature should continue, the amount of peculiar genetic features identified in SARS-CoV-2's genome does not rule out a possible gain-of-function origin, which should be therefore discussed in an open scientific debate.

## Conflict of interest

The authors declare no competing financial interest.

## Acknowledgment

We are grateful to the D.R.A.S.T.I.C. (Decentralised Radical Autonomous Search Team Investigating COVID-19) Twitter group for their investigative work in uncovering a significant number of previously unpublished facts about SARS-CoV-2 and its relative strains.